\documentclass[aps,pra,superscriptaddress,twocolumn]{revtex4-2}
\usepackage{graphicx,amsmath,amssymb,amsfonts,latexsym,color,dcolumn,bm}

\newcommand{\beq}{\begin{equation}}
\newcommand{\eeq}{\end{equation}}
\newcommand{\bea}{\begin{eqnarray}}
\newcommand{\eea}{\end{eqnarray}}

\providecommand{\bra}[1]{\langle #1 \rvert}
\providecommand{\ket}[1]{\lvert #1 \rangle}

\usepackage{amsfonts,amssymb}
\usepackage{dsfont}
\usepackage{physics}
\usepackage{tocvsec2}

\usepackage{appendix}

\usepackage{soul} 

\usepackage{tikz}

\usepackage{stmaryrd} 

\usepackage{amsmath}
\usepackage{bbold}
\usepackage{hyperref}
\hypersetup{
     colorlinks=true,      
    linkcolor=blue,        
    citecolor=blue,        
    filecolor=magenta, 
    urlcolor=black          
}

\usepackage{comment}

\begin{document}
\author{Mohamed Meguebel}
\email{mohamed.meguebel@telecom-paris.fr}
\affiliation{Telecom Paris, Institut Polytechnique de Paris, 19 Place Marguerite Perey, 91120 Palaiseau, France}
\author{Maxime Federico}
\affiliation{Telecom Paris, Institut Polytechnique de Paris, 19 Place Marguerite Perey, 91120 Palaiseau, France}
\affiliation{Laboratoire Interdisciplinaire Carnot de Bourgogne, UMR 6303 CNRS - Université de Bourgogne Franche-Comté, BP47870, 21078 Dijon, France}
\author{Louis Garbe}
\affiliation{Technical University of Munich, TUM School of Natural Sciences,
Physics Department, 85748 Garching, Germany}
\affiliation{Walther-Meißner-Institut, Bayerische Akademie der Wissenschaften, 85748 Garching, Germany and
Munich Center for Quantum Science and Technology (MCQST), 80799 Munich, Germany}
\author{Nadia Belabas}
\affiliation{Centre for Nanosciences and Nanotechnology, Université Paris-Saclay, 10 Bd Thomas Gobert, 91120 Palaiseau, France}
\author{Nicolas Fabre}
\email{nicolas.fabre@telecom-paris.fr}
\affiliation{Telecom Paris, Institut Polytechnique de Paris, 19 Place Marguerite Perey, 91120 Palaiseau, France}
\date{\today}
\begin{abstract}
We introduce a framework where light-matter transitions, rather than states, are the primary dynamical objects. Successive compositions of elementary transitions yield multiphoton processes with compact diagrammatic bookkeeping of resonant and off-resonant pathways. This approach enables transparent derivations of effective high-order Hamiltonians in the dispersive regime, foundational to quantum-information applications. Applied to the paradigmatic Jaynes-Cummings model, our framework reveals a photon-number-independent intrinsic Rabi frequency and persistent polaritonic hybridization in the dispersive regime, unifying resonant and dispersive limits. 
\end{abstract}
\vskip2pc

\title{Transitions as the Native Objects of Dispersive Light-Matter Dynamics}
\maketitle
\begin{figure}
    \centering
    \includegraphics[width=0.88\linewidth]{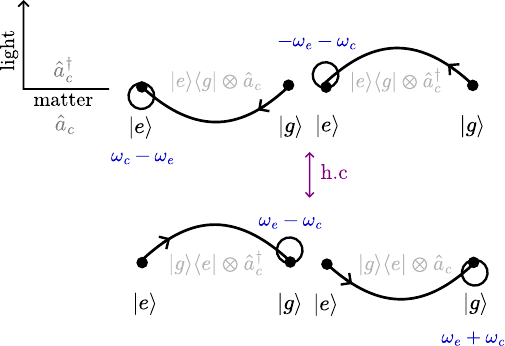}
    \caption{2D representation of JLM transition operators. The four 2D representations corresponding to the JLM transition operators $\ket{e}\bra{g}\otimes \hat a_c$, $\ket{e}\bra{g}\otimes \hat a_c^\dagger$, $\ket{g}\bra{e}\otimes \hat a_c^\dagger$ and $\ket{g}\bra{e}\otimes \hat a_c$ are associated to the detunings (in blue) $\omega_c-\omega_e$, $-\omega_e-\omega_c$, $\omega_e-\omega_c$ and $\omega_e+\omega_c$, respectively. Detunings are indicated by loops at the atomic states reached by the transitions and positioned below or above the matter axis depending on whether the photon is absorbed ($+\omega_c$) or emitted ($-\omega_c$). Hermitian conjugates -- associated with opposite detunings -- are obtained by mirror symmetry, \textit{i.e}, by reversing the transition direction (exchanging initial and final states) and replacing $\hat a_c\leftrightarrow \hat a_c^\dagger$. Detunings are directly inferred from the underlying atomic transitions together with the associated photon absorption or emission process.}
    \label{Figure JLM transition operators 2D representation}
\end{figure}
\begin{figure}
    \centering
    \includegraphics[width=0.65\linewidth]{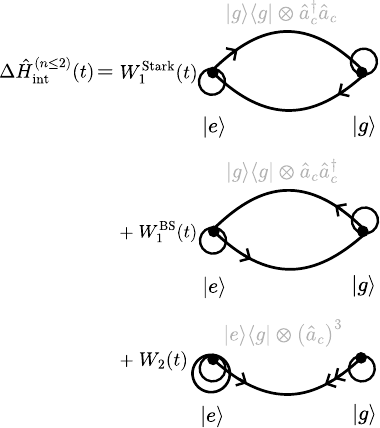}
    \caption{Formal representation of the interaction Hamiltonian correction $\Delta \hat H_{\text{int}}^{(n\leq 2)}(t) = \Delta \hat H_{\text{int}}^{(1)}(t)+\Delta \hat H_ {\text{int}}^{(2)}(t)$ up to second order. The time-dependent weights associated with $\ket{g}\bra{g}\otimes \hat a_c^\dagger \hat a_c$, $\ket{g}\bra{g}\otimes \hat a_c\hat a_c^\dagger$ and $\ket{e}\bra{g}\otimes (\hat a_c)^3$ are, $W_1^{\text{Stark}}(t)$, $W_1^{\text{BS}}(t)$ and $W_2(t)$, respectively, and the off-resonant diagrams mentioned in the text are sketched in the Supplementary Material but are discarded over a time large with respect to their associated detunings.}
    \label{Figure JLM diagrams 2LE perturbative expansion summary}
\end{figure}

Light-matter systems can operate in different interaction regimes, from weak coupling~\cite{Contextweakcouplingchang2006efficient} – where irreversible losses dominate – to strong coupling~\cite{Contextstrongcouplingungerer2024strong} – where coherent exchange of excitations governs the dynamics –, and extending to ultrastrong~\cite{ContextUltrastrongfrisk2019ultrastrong} and deep-strong coupling~\cite{Contextdeepstrongcasanova2010deep}, where the interaction strength becomes comparable to or exceeds the system frequencies. A canonical illustration of light-matter interaction in the closed-system limit is provided by the Jaynes–Cummings model~\cite{jaynes2005comparison,Quantumopticsjohn2014quantum,adiabaticeliminationatomicphysicslarson2021jaynes}. In the resonant or near-resonant regime, where the coupling $\lambda$ is parametrically larger than the detuning $\Delta$~\footnote{$\lambda$ and $\Delta$ are to be understood as formal parameters. In actual systems, $\lambda$ is the coupling term attached to a given light-matter transition while $\Delta$ is the corresponding detuning. In the present work, explicit detunings are written in the Rabi and Jaynes-Cummings models.}, an excitation is coherently exchanged between atom and cavity, giving rise to Rabi oscillations and dressed polaritonic eigenstates in which photonic and atomic degrees of freedom are hybridized. By contrast, when the detuning $\Delta$ is large compared with the coupling $\lambda$, the system enters the dispersive regime~\cite{Dispersiveregimezhu2013circuit}, where the eigenstates reduce to approximately separable light-matter product states, so that the polaritonic character effectively disappears and the interaction manifests through energy shifts rather than coherent excitation exchange. \\
The dispersive regime features rich high-order processes central to quantum technologies: it underpins superconducting-circuit architectures for quantum nondemolition measurements, lifetime enhancement, quantum gates, and collective effects~\cite{Dispersiveregimeblais2004cavity,Dispersiveregimeblais2007quantum,kakuyanagi2016observation}; enables two-photon Raman transitions in atom interferometry~\cite{Ramantwophotonlinskens1996two,Ramantwophotonbernard2022atom};  and has been implemented in solid-state platforms for two-photon entangling gate~\cite{Adiabaticeliminationmeguebel2025generation}. Furthermore, even paradigmatic cavity-atom systems exhibit highly nontrivial multiphoton interactions in the dispersive regime, including three-photon resonances as demonstrated by Ma and Law~\cite{Threephotoresonancenma2015three}. The dispersive regime thus offers a route to effective photon-photon interactions, motivating the development of methods to derive and systematically engineer higher-order multiphoton interactions. State-centric derivations include integro-differential techniques~\cite{Dispersiveregimepaulisch2014beyond}, resolvent-based projector~\cite{Dispersiveregimebrion2007adiabatic}, James’ effective-Hamiltonian methods~\cite{Jamesmethodjames2007effective,Jamesmethodgamel2010time,Jamesmethodshao2017generalized} and Schrieffer-Wolff transformation~\cite{SWbeyondRWAzueco2009qubit,SWbravyi2011schrieffer,Dispersiveregimezhu2013circuit,SWzhang2022quantum,SWayyash2025dispersive}. These methods rely on reference-frame~\cite{Dispersiveregimebrion2007adiabatic,Dispersiveregimepaulisch2014beyond} or abstract~\cite{SWbeyondRWAzueco2009qubit,SWbravyi2011schrieffer,Dispersiveregimezhu2013circuit,SWayyash2025dispersive} transformations, or on successive nested commutators~\cite{Jamesmethodjames2007effective,Jamesmethodgamel2010time,Jamesmethodshao2017generalized,SWbeyondRWAzueco2009qubit,SWbravyi2011schrieffer,Dispersiveregimezhu2013circuit,SWayyash2025dispersive} that are often difficult to track in high-order processes. \\
In our work, we place light-matter transition operators at the center of the dynamics, shifting the emphasis from states to transitions, from which both method and physical insights directly follow. \\
First, we show that higher-order effective interactions emerge as composite transitions built from successive concatenations of elementary operators, naturally organized within a compact diagrammatic, time-dependent perturbation theory; with a comprehensive treatment in a companion article~\cite{CompanionarticlePRA}. This approach bypasses Schrieffer-Wolff-type constructions and, for instance, recover the nontrivial three-photon resonance of Ma and Law~\cite{Threephotoresonancenma2015three} with reduced computational overload. \\ 
Second, our framework reveals a photon-number-independent Rabi frequency $\Omega = \sqrt{\Delta^2+4\lambda^2}$ in the Jaynes-Cummings model, emerging directly from the operator algebra. The intrinsic frequency $\Omega$ arises from a joint population operator that carries the polaritonic character across regimes, renormalizing the coupling on resonance and producing the Stark shift off resonance. We thus show that the dispersive regime is a reorganization of resonant polaritonic hybridization.\paragraph*{Rabi Hamiltonian} Consider a two-level atomic system with the excited state $\ket{e}$ of energy $\omega_e/2 > 0$ and the ground state $\ket{g}$ of energy $\omega_g = -\omega_e/2$. This two-level atomic system is placed within a cavity of frequency $\omega_c$. The Rabi Hamiltonian~\cite{RabimodelQuantxie2017quantum,JCmodellarson2021jaynes} can be expressed as ($\hbar =1$)
\begin{equation}
\label{Equation 2LE microscopic Hamiltonian}
\hat H_{\text{R}} = \frac{\omega_e}{2}\hat \sigma_z+\omega_c\hat n_c  + \lambda \hat \sigma_x\otimes(\hat a_c+\hat a_c^\dagger),
\end{equation}
with $\hat \sigma_z = \ket{e}\bra{e}-\ket{g}\bra{g}$ and $\hat \sigma_x = \hat \sigma_+ + \hat \sigma_-$, where $\hat \sigma_+ = \ket{e}\bra{g}$ and $\hat \sigma_- = \ket{g}\bra{e}$ are the raising and lowering atomic operators, respectively. $\hat a_c$ and $\hat a_c^\dagger$ are, respectively, the bosonic annihilation and creation operators, and $\hat n_c = \hat a_c^\dagger \hat a_c$ the photon-number operator. The eigenstates of the free Hamiltonian $\hat H_{\text{free}} = \frac{\omega_e}{2}\hat \sigma_z+\omega_c\hat n_c$ are the bare states $\ket{k,n}$, with $n$ the number of photons and $k=e,g$. Their eigenvalues are, respectively, $n\omega_c \pm \omega_{e}/2$ and correspond to the energies of the uncoupled light-matter system. In a state-centric description, the $\ket{k,n}$ are the zeroth-order eigenobject prior to interaction. The interaction Hamiltonian 
\begin{equation}
\label{Equation Rabi interaction Hamiltonian}
    \hat H_{\text{int}} =  \lambda \hat \sigma_x\otimes(\hat a_c+\hat a_c^\dagger),
\end{equation}
as opposed to the Jaynes-Cummings~\cite{Quantumopticsjohn2014quantum} interaction Hamiltonian, accounts for processes both within and beyond the rotating-wave approximation (RWA), with the corresponding operators $\ket{e}\bra{g}\otimes \hat a_c + \text{h.c}$ and $\ket{e}\bra{g}\otimes \hat a_c^\dagger + \text{h.c}$, respectively. \paragraph*{Transition representation} Rather than describing the dynamics at the level of states, we formulate the dynamics in terms of the interaction processes. We define the transition-centric zeroth-order eigenobjects -- \textit{prior to interaction} -- to be \textit{joint light–matter} (JLM) transition operators $\hat \xi_i$s defined as the elementary one-photon operator components of the interaction Hamiltonian: $\ket{e}\bra{g}\otimes \hat a_c$, $\ket{e}\bra{g}\otimes \hat a_c^\dagger$ and their Hermitian conjugates. This underscores a fundamental distinction between state-centric and transition-centric frameworks: uncoupled basis states $\ket{k,n}$ in Hilbert space contain no information about the light-matter interaction, whereas JLM transition operators in Liouville space~\cite{Liouvillespacegyamfi2020fundamentals} -- the space of operators acting on the Hilbert space -- are the elementary building blocks of the interaction Hamiltonian Eq.~\eqref{Equation Rabi interaction Hamiltonian}. In analogy with the state-centric description, one may associate eigenvalues with JLM transition operators by examining their transformation under a superoperator, \textit{i.e.}, an operator acting on operators. The JLM transition operators are eigenoperators~\cite{DoublesidedFeynmandiagramsmukamel1995principles,Generalopenquantumsystembreuer2002theory} of the free Liouvillian $\mathcal L_{\text{free}}$ acting on any JLM transition operator $\hat \xi_i$ as $\mathcal L_{\text{free}}\hat \xi_i = [\hat \xi_i, \hat H_{\text{free}}] = \Delta_{\hat \xi_i} \hat \xi_i$; and whose eigenvalues $\Delta_{\hat \xi_i}$ are not energies but detunings. Transitions are thus primary objects characterized by their detunings eigenvalues. For a given detuning, infinitely many energy assignments are compatible with the same transition, making detunings the irreducible spectral attributes of light-matter dynamics.\\
JLM transition operators can be represented graphically, much like atomic states are depicted in standard ladder diagrams. In Fig.~\ref{Figure JLM transition operators 2D representation}, we adopt a 2D representation in which the light dynamics is tracked along one axis and the matter dynamics along the other.
This representation exploits the interaction Hamiltonian formulation in terms of photon creation and annihilation operators, allowing transitions to be visualized directly as elementary creation and annihilation events rather than through photon-number states.\paragraph*{Transition-centered perturbative expansion}
A time-independent JLM transition operator $\hat \xi_i$ from $\hat H_{\text{int}}$ (see Eq.~\eqref{Equation Rabi interaction Hamiltonian}) becomes time-dependent in the Heisenberg-picture~\cite{Generalsakurai2020modern}, $\hat \xi_i \rightarrow \hat \xi_i(t)$, and evolves for $t>0$ as $d\hat \xi_i(t)/dt = -i\mathcal L_\text{R}\left(\hat \xi_i(t)\right)$ under the adjoint Liouvillian~\cite{Generalopenquantumsystembreuer2002theory,Liouvillespacegyamfi2020fundamentals} $\mathcal L_{\text{R}}(*) = [*,\hat H_{\text{R}}] = \mathcal L_{\text{free}}(*)+\mathcal L_{\text{int}}(*)$, where $\mathcal L_{\text{int}}(*) = [*,\hat H_{\text{int}}]$ is the interaction Liouvillian. Each JLM transition operator admits a perturbative expansion in successive orders of the coupling term $\lambda$, \textit{e.g.} $\hat \xi_i(t) = \sum_{n=0}^{+\infty}\hat \xi_i^{(n)}(t)$, where $\hat \xi_i^{(n)}(t)$ is proportional to $\lambda^n$.
The full expansion is captured by the time-dependent superoperator $\mathcal L_{\text{int}}(\tau)=e^{i\mathcal L_{\text{free}}\tau}\mathcal L_{\text{int}}e^{-i\mathcal L_{\text{free}}\tau}$, whose action concatenates elementary transitions. Each application generates a higher-order transition that remains an eigenoperator of $\mathcal L_{\text{free}}$, with eigenvalue given by the sum of detunings of its constituents. For instance, at first order $\hat \xi_i^{(1)}(t)= -i\lambda\sum_{j\neq i}\int_{0}^t d\tau e^{-i(\delta_i+\delta_j)(t-\tau)}e^{-i\delta_i\tau} [\hat \xi_i, \hat \xi_j]$, where $\delta_i$ and $\delta_j$ are the detuning eigenvalues associated with the time-independent operators $\hat \xi_i$ and $\hat \xi_j$ from $\hat H_{\text{int}}$ (see Eq.~\eqref{Equation Rabi interaction Hamiltonian}), respectively, and $\delta_i+\delta_j$ the cumulative detuning corresponding to the first-order operator $[\hat \xi_i,\hat \xi_j]$. As detailed in the Supplementary Material and in the companion article~\cite{CompanionarticlePRA}, higher-order terms can systematically be derived upon toggling to Laplace space.\paragraph*{Diagrammatic structure} The concatenation structure maps directly onto a diagrammatic representation. Indeed, each JLM transition from $\hat H_{\text{int}}$ (see Eq.~\eqref{Equation Rabi interaction Hamiltonian}) is an elementary block, and their successive composition produces higher-order transitions whose detunings, operator ordering and selection rules are encoded visually. Counter-rotating processes enter on the same footing as co-rotating contributions, appearing as additional pathways, thereby unifying RWA and non-RWA dispersive regimes~\cite{Dispersiveregimezueco2009qubit,Dispersiveregimemuller2020dissipative,SWayyash2025dispersive}. Crucially, while state-centric perturbation theory navigates a fixed eigenstate space $\{\ket{k,n}\}_{k=e,g}$, transition-based perturbation theory generates new transitions as dynamical objects whose frequency content and origin are encoded directly in their diagrammatic construction. The transition-centric framework thus provides a systematic and transparent route to effective Hamiltonians at arbitrary perturbative order, which we now derive.\paragraph*{High-order effective Hamiltonians}
The derivation of higher-order effective Hamiltonians -- with a systematic workflow provided in the End Matter -- reduces to performing the perturbative expansion of the elementary JLM transition operators already present in the first-order interaction Hamiltonian. At order $n$, reinjecting every perturbatively expanded transition $\hat \xi_i^{(n)}(t)$ yields a correction $\Delta \hat H_{\text{int}}^{(n)}(t)$ to the interaction Hamiltonian, as represented in Fig.~\ref{Figure JLM diagrams 2LE perturbative expansion summary}, consisting of $n$th-order JLM transition operators with weights $W_n(t)$ in $\Delta \hat H_{\text{int}}^{(n)}(t)$ growing as $\lambda^{n+1}$, which we compute as 
\begin{align}
\label{Equation time-dependent weight in the first, second and third order}
W_0(t) &= \lambda e^{-i\delta_i t} \\
\label{Equation first-order time-weight}
W_1(t) &\approx \frac{\lambda^2}{2}e^{-i(\delta_i+\delta_j)t}\left[\frac{1}{\delta_i}-\frac{1}{\delta_j}\right] \\
\label{Equation second-order time-weight}
W_2(t) &\approx \frac{\lambda^3}{3}e^{-i(\delta_i+\delta_j+\delta_k)t} \notag \\
&\quad \times \left[\frac{1}{\delta_i(\delta_i+\delta_j)}
+\frac{1}{\delta_k(\delta_k+\delta_j)}
-\frac{1}{\delta_i\delta_k} \right],
\end{align}
where $\delta_i$, $\delta_j$, and $\delta_k$ denote the one-photon detunings associated with the first, second, and third elementary JLM transition operators entering the construction of an $n$th-order composite transition. 
Computation details are provided in the Supplementary Material and in the companion article~\cite{CompanionarticlePRA}, wherein waveguide QED systems are further discussed. Each $n$th-order JLM transition operator carries a weight $W_n(t)$ given by a product of $n$ coupling-to-detuning ratios, thereby making the dispersive scaling explicit -- namely, that the perturbative expansion holds when $\lambda$ is small compared to the detunings -- without requiring any prior reference-frame transformation. We stress that the expressions Eq.~\eqref{Equation first-order time-weight} and Eq.~\eqref{Equation second-order time-weight} are straightforward to handle. They are indeed computed directly from reading off the detunings associated with the elementary building blocks of a given multiphoton JLM transition operator. The expressions for $W_1(t)$ and $W_2(t)$ are obtained by eliminating fast-varying contributions at large times compared to the inverse off-resonant detunings~following the logic of adiabatic elimination~\cite{Jamesmethodjames2007effective,Adiabaticeliminationbeyondpaulisch2014beyond}. The expression for $W_2(t)$ is valid provided that no partial sum of detunings vanishes, specifically $\delta_i \neq -\delta_j$ and $\delta_k \neq -\delta_j$. These conditions exclude cumulative-detuning degeneracies that may arise at higher perturbative order and require a separate regularization procedure. Such cases are discussed in detail in the companion article~\cite{CompanionarticlePRA}.\paragraph*{Three-photon resonance} We illustrate the transition-centric perturbation theory by retrieving the three-photon effective Hamiltonian obtained by Law and Ma~\cite{Threephotoresonancenma2015three} where $\omega_c \approx \omega_e/3$ with significantly reduced computational overload. Order-by-order details are given in the Supplementary Material. Consider the JLM transition operator $\left(\ket{e}\bra{g}\otimes \hat a_c \right)(t) = \sum_{n=0}^{2}\left( \ket{e}\bra{g}\otimes \hat a_c\right)^{(n)}(t)$ perturbative expansion up to second order. Its zeroth-order contribution is given by $e^{-i(\omega_c-\omega_e)t}\ket{e}\bra{g}\otimes \hat a_c$, which oscillates at a detuning $ \delta_i= \omega_c-\omega_e \simeq -2\omega_c$. This rapid oscillation causes the operator to average to zero over a time $T \gg 1/(2\omega_c)$, and the term can therefore be neglected. The reasoning is identical for $\left(\ket{e}\bra{g}\otimes \hat a_c^\dagger\right)(t)$ and the Hermitian conjugates. Second, sketching the first-order JLM diagrams show two different diagram topologies: (i) The first topology corresponds to open diagrams where the atomic state goes back to the initial one with two absorbed photons or two emitted photons. These diagrams amount to first-order JLM transition operators of the form $\ket{i}\bra{i}\otimes \hat a_c \hat a_c$ or $\ket{i}\bra{i}\otimes \hat a_c^\dagger \hat a_c^\dagger$, with $i = e,g$. These operators accumulate a phase $\pm 2\omega_c t$ and are subsequently disregarded over a time $T\gg 1/(2\omega_c)$; (ii) In Fig.~\ref{Figure JLM diagrams 2LE perturbative expansion summary}, we show the second topological structure consisting in loop JLM diagrams, in which the diagram closes onto the initial atomic state. The associated first-order JLM transition operators therefore oscillate at zero frequency and do not average out to zero. The effective-Hamiltonian correction arising from the second-topology diagram can now be calculated using Eq.~\eqref{Equation first-order time-weight}
\begin{equation}
\begin{split}
\label{Equation first-order correction to the Rabi interaction Hamiltonian}
    \Delta \hat H^{(1)}_{\text{int}} &= \left(\frac{\lambda^2}{\omega_e-\omega_c}+\frac{\lambda^2}{\omega_e+\omega_c}\right)\left(\hat \sigma_z \hat n_c+\frac{\hat \sigma_z}{2} \right),
\end{split}
\end{equation}
up a global energy constant.  When the virtual pathways involve only co-rotating processes, the renormalization appears as a quantum Stark shift~\cite{QuantumACStarkshiftblais2004cavity}, embodied by the $\lambda^2/(\omega_e-\omega_c)$ term in Eq.~\eqref{Equation first-order correction to the Rabi interaction Hamiltonian}. In contrast, pathways containing counter-rotating events produce the Bloch-Siegert (BS) shift~\cite{BlochSiegertbloch1940magnetic,BlochSiegertforn2010observation}, associated with the term proportional to $\lambda^2/(\omega_e+\omega_c)$ in Eq.~\eqref{Equation first-order correction to the Rabi interaction Hamiltonian}.
\begin{figure}
    \centering
    \includegraphics[width=0.7\linewidth]{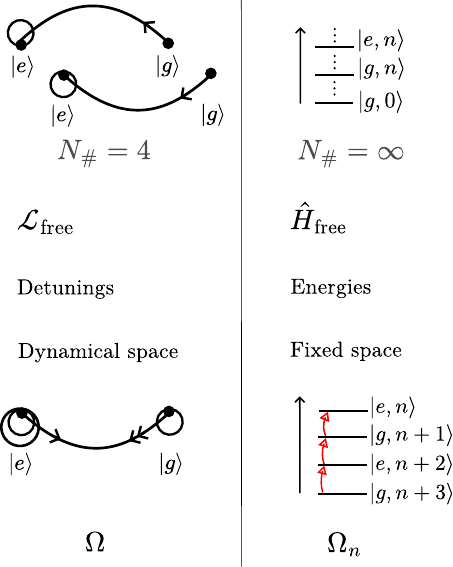}
    \caption{Comparison between the state (bottom) and transition (top) perspective. In the state-centric view, the zeroth-order objects are the states $\ket{k,n}_{k=e,g}$ (infinitely many in the Rabi Hamiltonian), with energies set by the free Hamiltonian. The interaction Hamiltonian induces transitions within this fixed state space. In the Jaynes-Cummings model, dynamics occur in subspaces $\{\ket{e,n},\ket{g,n+1}\}_{n\geq 0}$ with photon-number-dependent Rabi frequency $\Omega_n = \sqrt{\Delta^2+4\lambda^2(n+1)}$, yielding polaritonic dressed states. In the transition-centric view, the zeroth-order objects are the JLM transition operators (four in the Rabi Hamiltonian), whose eigenvalues are given by the free Liouvillian and correspond to detunings. The interaction Liouvillian couples transitions, dynamically extending the space. In the Jaynes-Cummings model, the operator subspace $\mathcal S = \{\hat n_c,\ \hat \sigma_z/2 + \hat \sigma_z \hat n_c,\ \hat \sigma_+\hat a_c, \hat \sigma_-\hat a_c^\dagger\}$ exhibits Rabi oscillations with photon-number-independent frequency $\Omega = \sqrt{\Delta^2+4\lambda^2}$, arising from the polaritonic joint population term $\hat \sigma_z \hat n_c$.}
    \label{Figure state vs transition  representations}
\end{figure}
Lastly, let us scrutinize second-order JLM diagrams associated with the three-photon resonance. One can immediately discard most of the JLM diagrams. Indeed, all diagrams are going to be of the form $\ket{e}\bra{g}\otimes \dots$ (and Hermitian conjugate) where $\dots$ represents combinations of three bosonic operators. Due to the three-photon resonance condition $\omega_c\approx \omega_e/3$, only the term in $\ket{e}\bra{g}\otimes \hat a_c \hat a_c \hat a_c$, showcased in Fig.~\ref{Figure JLM diagrams 2LE perturbative expansion summary}, and its Hermitian conjugate are going to result in resonant contributions. The other diagrams are dynamically suppressed over timescales long compared to the one-photon detunings. Therefore, the second-order correction to the interaction Hamiltonian reads 
\begin{equation}
\label{Equation second-order correction to the interaction Hamiltonian}
    \Delta \hat H^{(2)}_{\text{int}} = -\frac{9\lambda^3}{4\omega_e^2}\hat \sigma_+\otimes (\hat a_c)^3 + \text{h.c},
\end{equation}
where we used $W_2(t)$ (see Eq.~\eqref{Equation second-order time-weight}) and $\omega_e \approx 3\omega_c$. In the $\{\ket{e,0}, \ket{g,3}\}$ state subspace, $\Delta \hat H^{(1)}_{\text{int}}$ (see Eq.~\eqref{Equation first-order correction to the Rabi interaction Hamiltonian}) and $\Delta \hat H^{(2)}_{\text{int}}$ (see Eq.~\eqref{Equation second-order correction to the interaction Hamiltonian}) can be cast as 
\begin{equation}
\label{Equation first-order shift to the Hamiltonian two-photon resonance}
\begin{split}
    \Delta \hat H^{(1)}_{\text{int}} &= \frac{3\lambda^2}{2\omega_e}\ket{e,0}\bra{e,0}-\left(\frac{9\lambda^2}{2\omega_e}+\frac{\lambda^2}{\omega_c}\right)\ket{g,3}\bra{g,3}
\end{split}
\end{equation}
and 
\begin{equation}
\label{Equation second-order shift to the Hamiltonian three-photon resonance}
    \Delta \hat H^{(2)}_{\text{int}} = -\frac{9\sqrt{6}\lambda^3}{4\omega_e^2}\ket{e,0}\bra{g,3}+\text{h.c}.
\end{equation}
Eq.~\eqref{Equation first-order shift to the Hamiltonian two-photon resonance} and Eq.~\eqref{Equation second-order shift to the Hamiltonian three-photon resonance} can further be adapted to the basis $\{\ket{e,n}, \ket{g,n+3}\}$ with $n \geq 0$. Eq.~\eqref{Equation first-order shift to the Hamiltonian two-photon resonance} and Eq.~\eqref{Equation second-order shift to the Hamiltonian three-photon resonance} coincide exactly with the effective terms derived by Law and Ma in~\cite{Threephotoresonancenma2015three}. \\
Our JLM transition-operator diagrammatic approach offers several key advantages: (A1) It operates directly on transition operators, incorporating all states accessible through these transitions without \textit{a priori} Hilber-space truncation. (A2) It reduces computational overhead compared with state-centric approaches, as it requires neither Hamiltonian diagonalization, problem-dependent generators, nor predefined relevant and irrelevant subspace, while preserving perturbative traceability through diagrammatic bookkeeping. (A3) Resonant and non-resonant pathways are identified directly at the transition level. In particular, whereas counter-rotating pathways underlying the three-photon resonance remain implicit in previous derivations~\cite{Threephotoresonancenma2015three}, the JLM diagrams in Fig.~\ref{Figure JLM diagrams 2LE perturbative expansion summary} make them explicit, enabling a systematic treatment of higher-order interactions. Fig.~\ref{Figure state vs transition representations} presents a schematic comparison of transition-centric and state-centric descriptions of light-matter dynamics. The transition-centric framework not only offers practical advantages but also uncovers a fundamental physical consequence: the persistent polaritonic hybridization of light and matter across both the resonant and dispersive regime.  \paragraph*{Intrinsic Rabi frequency} Consider now the simpler Jaynes-Cummings Hamiltonian 
\begin{equation}
\label{Equation Jaynes-Cummings Hamiltonian}
\hat H_{\text{JC}} = \frac{\omega_e}{2}\hat \sigma_z+\omega_c\hat n_c+\lambda \left(\hat \sigma_+\hat a_c+\hat \sigma_-\hat a_c^\dagger \right);
\end{equation}
by dropping the counter-rotating transitions from the Rabi Hamiltonian Eq.~\eqref{Equation 2LE microscopic Hamiltonian}. In the state-centric description, the subspaces $\{\ket{e,n},\ket{g,n+1}\}_{n\geq 0}$ of the Jaynes-Cummings Hamiltonian are closed, reducing the problem to a $2\times2$ diagonalization with vacuum Rabi-split eigenvalues and dressed-state polariton eigenvectors, both characterized by the Rabi frequency $\Omega_n = \sqrt{\Delta^2+4\lambda^2(n+1)}$. In an operator-centric description, one may consider the operator subspace $\mathcal S' = \{ \hat n_c, \hat \sigma_z,\hat \sigma_+\hat a_c, \hat \sigma_-\hat a_c^\dagger\}$ constitutive of the Jaynes-Cummings Hamiltonian Eq.~\eqref{Equation Jaynes-Cummings Hamiltonian}. The operators $\hat \sigma_+\hat a_c$ and $\hat \sigma_-\hat a_c^\dagger$ correspond to the JLM transition operators discussed earlier, forming the \textit{transition sector} $\{\hat \sigma_+\hat a_c, \hat \sigma_-\hat a_c^\dagger\}$, while the operators $\hat n_c$ and $\hat \sigma_z$ define the \textit{disjoint population} sector $\{\hat \sigma_z, \hat n_c\}$ associated with disjoint light-matter self-transitions. We demonstrate in the Supplementary Material that the time-evolution characteristic frequency within $\mathcal S'$ is a photon-number-independent frequency $\Omega' = \sqrt{\Delta^2+2\lambda^2}$, arising solely from the Jaynes-Cummings Hamiltonian, that is to say the light-matter dynamics. We therefore term $\Omega'$ the \textit{intrinsic Rabi frequency}, in contrast to the photon-number-dependent $\Omega_n$. However, this value does not reproduce $\Omega_n$ when the Hilbert space is restricted to $\{\ket{e,n},\ket{g,n+1}\}_{n\geq 0}$, which rescales the coupling term $\lambda  \rightarrow \lambda \sqrt{n+1}$ through $\hat \sigma_+\hat a_c \ket{g,n+1} = \sqrt{n+1}\ket{e,n}$ and $\hat \sigma_-\hat a_c^\dagger \ket{e,n} = \sqrt{n+1}\ket{g,n+1}$. The reason is that, as previously discussed, the operator subspace $\mathcal S'$ is not closed under the adjoint Liouvillian $\mathcal L_{\text{JC}}(*) = -i[*,\hat H_{\text{JC}}]$. The interaction Hamiltonian indeed mixes transitions and generates additional terms. Specifically, it produces an additional term $\hat \sigma_z \hat n_c$ which we identify as a \textit{polaritonic joint population} term, which has no state-centric counterpart. Dressing the disjoint population operator $\hat \sigma_z$ by the polaritonic correction $\hat \sigma_z \rightarrow \hat \pi = \hat \sigma_z/2 + \hat \sigma_z \hat n_c$ and retaining only this lowest nontrivial contribution, the reduced set $\mathcal S = \{\hat n_c, \hat \pi, \hat \sigma_+\hat a_c, \hat \sigma_-\hat a_c^\dagger\}$ evolves with a characteristic frequency $\Omega = \sqrt{\Delta^2 + 4\lambda^2}$. The photon-number-dependent Rabi frequency $\Omega_n$ can be then deduced from the Rabi frequency $\Omega$ upon restricting the Jaynes-Cummings Hamiltonian Eq.~\eqref{Equation Jaynes-Cummings Hamiltonian} to the state subspace $\{\ket{e,n},\ket{g,n+1}\}_{n\geq 0}$. This shows that the transition-centric Rabi frequency $\Omega$ is the primary, intrinsic frequency of the light-matter interaction, while the state-centric, photon-number-dependent $\Omega_n$ arises only after projecting onto a fixed photon-number manifold. The intrinsic Rabi frequency $\Omega$ moreover coincides with the vacuum Rabi frequency of the Jaynes-Cummings manifold $\{\ket{e,0}, \ket{g,1}\}_{n=0}$, supporting its interpretation as a fundamental dynamical frequency of the light-matter interaction, independent of the photonic state.\paragraph*{Polaritonic joint population} The operator $\hat \sigma_z\hat n_c$ -- neither purely light nor purely matter -- encodes the polaritonic nature of the interaction and captures the dominant light-matter dynamics despite the Liouvillian generating additional operators beyond the reduced space $\mathcal S$.  At resonance, it renormalizes the coupling term $2\lambda^2 \rightarrow 4\lambda^2$, thus restoring the intrinsic Rabi frequency $\Omega = \sqrt{\Delta^2 + 4\lambda^2}$ consistent with $\Omega_n$. In the dispersive regime, the two forms $\Omega'$ and $\Omega$ of the intrinsic Rabi frequency coincide with $|\Delta|$, and the joint population term $\hat \sigma_z\hat n_c$ captures the residual hybrid nature of the interaction with a Stark shift, as observed in Eq.~\eqref{Equation first-order correction to the Rabi interaction Hamiltonian}. Placing light-matter operators at center stage thus provides a compact operator-space explanation linking resonant coupling renormalization to dispersive detuning shifts as two manifestations of the same polaritonic interaction.\paragraph*{Rabi oscillations in the dispersive regime} The intrinsic Rabi frequency $\Omega$ governs oscillations of the operator subspace in both the population sector $\{\hat n_c, \hat \pi\}$ and the transition sector $\{\hat \sigma_+ \hat a_c, \hat \sigma_-\hat a_c^\dagger\}$. Indeed, we show in the Supplementary Material that in the resonant regime, the symmetric transition operator $\hat \sigma_+\hat a_c + \hat \sigma_-\hat a_c^\dagger$ is frozen while the oscillatory subspace is spanned by the population operators and the antisymmetric transition operator $\hat \sigma_+\hat a_c - \hat \sigma_-\hat a_c^\dagger$. In contrast, in the dispersive regime, the population operators freeze while the transition sector carries the Rabi oscillations. This interpretation maps directly onto the Bloch-vector description of the density matrix: the Bloch vector always precesses, but the precession axis -- and hence which components show apparent oscillations -- changes with the interaction regime; resonant or dispersive. \\ 
\paragraph*{Conclusion and outlook} We establish light-matter operators as the native objects of light-matter dynamics. This perspective provides a ready-to-use diagrammatic perturbation theory in which higher-order effective interactions emerge as concatenated elementary transitions, with operator ordering, detunings, and selection rules encoded directly in the diagrams, rendering virtual pathways explicit. Our framework also reveals the unified hybrid nature of resonant and dispersive light-matter interactions: the same transition-level structure governs polaritonic hybridization in both regimes, manifesting as coupling renormalization on resonance and as detuning renormalization in the dispersive limit. This operator-centric framework applies uniformly to discrete-mode spectra and frequency continua, placing cavity and waveguide QED on the same conceptual footing, and thus provides a practical tool for engineering high-order squeezing~\cite{Highsqueezingsutherland2021universal,buazuavan2024squeezing} operations.  Moreover, by elucidating the inherent structure of light-matter interactions and hybridization across different regimes, the framework paves the way for a deeper understanding of correlated light-matter systems and vacuum phenomena~\cite{Vacuumeffectguerin2019collective,VacuumPhysRevA.110.063703}. In that respect, perspectives include identifying experimental settings where the intrinsic Rabi frequency captures physics beyond what the photon-number-dependent Rabi frequency alone provides.\paragraph*{Acknowledgments}
M. Meguebel acknowledges support from the Program QuanTEdu-France n° ANR-22-CMAS-0001 France 2030. L. Garbe acknowledges funding from the Munich Quantum Valley, which is supported by the Bavarian state government with funds from the Hightech Agenda Bayern Plus.
\bibliography{ref_JLM_dia}
\section*{End Matter}
\paragraph*{JLM diagrams general construction rules and workflow} We outline a practical set of rules and a workflow for constructing JLM diagrams, enabling the derivation of effective high-order Hamiltonians in the dispersive regime up to second order ($n \leq 2$). The procedure is adapted from the companion article~\cite{CompanionarticlePRA}, where additional details can be found.
\begin{itemize}
\item (S1) Start from the interaction Hamiltonian and identify its fundamental building blocks, namely the zeroth-order JLM transition operators.
\item (S2) Construct the $n$th-order JLM diagrams by concatenating $n+1$ zeroth-order transition operators.
\item (S3) Compute the weight of each $n$th-order operator (up to $n=2$) using the weights $W_0(t)$, $W_1(t)$ and $W_2(t)$ defined in the main text.
\item (S4) Sum all $n$th-order JLM transition operators, each multiplied by its corresponding weight $W_n(t)$, to obtain $\Delta \hat H_{\text{int}}^{(n)}(t)$.
\item (S5) Finally, perform the adiabatic elimination by discarding off-resonant contributions over a timescale $T$ that is large compared to the inverse of their associated detunings.
\end{itemize}
\end{document}


\author{Mohamed Meguebel\footnote{mohamed.meguebel@telecom-paris.fr}}
\affiliation{Telecom Paris, Institut Polytechnique de Paris, 19 Place Marguerite Perey, 91120 Palaiseau, France}
\author{Maxime Federico}
\affiliation{Telecom Paris, Institut Polytechnique de Paris, 19 Place Marguerite Perey, 91120 Palaiseau, France}
\affiliation{Laboratoire Interdisciplinaire Carnot de Bourgogne, UMR 6303 CNRS - Université de Bourgogne Franche-Comté, BP47870, 21078 Dijon, France}
\author{Louis Garbe}
\affiliation{Technical University of Munich, TUM School of Natural Sciences,
Physics Department, 85748 Garching, Germany}
\affiliation{Walther-Meißner-Institut, Bayerische Akademie der Wissenschaften, 85748 Garching, Germany and
Munich Center for Quantum Science and Technology (MCQST), 80799 Munich, Germany}
\author{Nadia Belabas}
\affiliation{Centre for Nanosciences and Nanotechnology, Université Paris-Saclay, 10 Bd Thomas Gobert, 91120 Palaiseau, France}
\author{Nicolas Fabre\footnote{nicolas.fabre@telecom-paris.fr}}
\affiliation{Telecom Paris, Institut Polytechnique de Paris, 19 Place Marguerite Perey, 91120 Palaiseau, France}
\date{\today}
\begin{abstract}
\end{abstract}
\pacs{}
\vskip2pc

\title{Transitions as the Native Objects of Dispersive Light-Matter Dynamics -- Supplementary Material}
\maketitle
\onecolumngrid

\subsection*{Joint light-matter transition operator weight calculation}
We derive the weight $W_n(t)$ attached to a given $n$th-order JLM operator up to second order. The computation builds on the weight $w_n(t)$ of an individual JLM diagram constructed from a given zeroth-order JLM operator. The weight $w_n(t)$ is not to be confused with $W_n(t)$ which corresponds to the total weight of a given $n$th-order operator in $\Delta \hat H_{\text{int}}^{(n)}(t)$. Indeed, different unphysical perturbation time-orderings or choices of zeroth-order JLM operators in the expansion can produce the same $n$th order JLM transition operator, but with different weights $w_n(t)$.
We account for this by defining $W_n(t)$ as the average of these various contributions $w_n(t)$. The derivation is carried through in the companion article~\cite{CompanionarticlePRA} in a more general setting. In particular, in~\cite{CompanionarticlePRA} we extend the formalism to frequency-dependent JLM transition operators and couplings, a feature that is essential in waveguide QED. In what follows, the coupling terms in the interaction Hamiltonian $\hat H_{\text{int}}$ are considered identical and equal to $\lambda$. \\
\subsubsection*{Weight of a JLM transition diagram} 
A time-independent JLM transition operator $\hat \xi$ drawn from the interaction Hamiltonian $\hat H_{\text{int}}$ becomes time-dependent in the Heisenberg picture -- that is $\hat \xi \rightarrow \hat \xi(t)$ --, and evolves under the Liouvillian $\mathcal L = \mathcal L_{\text{free}}+\mathcal L_{\text{int}}$, with $\mathcal L_{\text{free}}(\hat \xi) = [\hat \xi,\hat H_{\text{free}}]$ and $\mathcal L_{\text{int}}(\hat \xi) = [\hat \xi, \hat H_{\text{int}}]$. Consider the gauge transformation $\hat \xi(t) \rightarrow e^{-i\mathcal L_{\text{free}}t}\hat \xi^{(I)}(t)$, where $\hat \xi^{(I)}(t)$ will denote the interaction-picture JLM transition operator. The time-evolution of $\hat \xi^{(I)}(t)$ is given by the time-evolution superoperator $\mathcal U_I(t)$ as $\hat \xi^{(I)}(t) = \mathcal U_I(t)\hat \xi^{(I)}(0)$, with 
\begin{equation}
     \mathcal{U}_I(t) = \mathcal T_{\leftarrow} \exp\left(-i\int_{0}^td\tau\;\mathcal L_{\text{int}}(\tau) \right),
\end{equation}
where $\mathcal T_{\leftarrow}$ is the time-ordering operator. Expanding $\mathcal{U}_I(t,t_0)$ and $\hat \xi(t)$ in powers of $\lambda$, in particular $\hat \xi(t) = \sum_{n=0}^{+\infty} \hat \xi^{(n)}(t)$, terms are identified order-wise 
\begin{equation}
    \begin{split}
        \hat \xi^{(0)}(t) &= e^{-i\mathcal L_{\text{free}}t}\hat \xi \\ 
    \hat \xi^{(1)}(t) &= -i\int_{0}^td\tau\; e^{-i\mathcal L_{\text{free}}t}\mathcal L_{\text{int}}e^{-i\mathcal L_{\text{free}}\tau}\hat \xi \\ 
    \hat \xi^{(2)}(t) &= (-i)^2\int_{0}^t\int_{0}^\tau d\tau d\tau'\; e^{i\mathcal L_{\text{free}}\tau}\mathcal L_{\text{int}}e^{-i\mathcal L_{\text{free}}(\tau-\tau')}\mathcal L_{\text{int}}e^{-i\mathcal L_{\text{free}}\tau'}\hat \xi \\ 
    &\;\;\vdots,
    \end{split}
\end{equation}
where we went back to the Heisenberg picture upon applying $e^{-i\mathcal L_{\text{free}}t}$ and used the fact that at the initial time $t=0$, all pictures are equivalent. A $n$th-order JLM diagram has weight $w_n(t)$ in the $n$th-order correction $\Delta \hat H_{\text{int}}^{(n)}(t)$ to the interaction Hamiltonian upon reinjecting the $\hat \xi^{(n)}(t)$ in the interaction Hamiltonian. $w_n(t)$ may be expressed as 
\begin{equation}
    w_n(t) = \lambda^{n+1}v_n(t),
\end{equation}
with 
\begin{equation}
    v_n(t) = (-i)^n(-1)^{N_L}\int_{0}^{t}d\tau_{n} \dots \int_{0}^{\tau_2}d\tau_1 e^{-i\Delta_n(t-\tau_{n})}\dots e^{-i\Delta_0\tau_1},
\end{equation}
where the $\Delta_i$s are the detunings involved at the $i$th order and where the factor $(-1)^{N_L}$ accounts for the $N_L$ operators acting on the left of the zeroth-order JLM transition operator and arises from the commutator with $\hat H_{\text{int}}$ in the interaction Liouvillian. The detunings are cumulative, \textit{i.e.} $\Delta_i = \sum_{l=0}^i \delta_l$, where $\delta_l$ is the detuning of the JLM transition driving the $l$th-order perturbation. Accordingly, the $i$th-order detuning depends on those of all preceding orders. In line with the companion article~\cite{CompanionarticlePRA}, we associate to each individual detuning $\delta_l$ a regularization term $-i\theta_l$ with $\theta_l>0$. At the end of the computation, we take the limit $\theta_l \rightarrow 0$, ensuring that no additional physical terms are introduced. This regularization is required when treating continua, for instance in waveguide QED, and is also well suited to handle detuning degeneracies, as discussed next. The time-dependent term $v_n(t)$ -- which amounts to nested convolution products -- can be handled by taking the Laplace transform 
\begin{equation}
    F_n(t) = \mathcal L_a[v_n(t)] = (-i)^n\prod_{k=0}^n\frac{1}{s+i\Delta_k+\Theta_k},
\end{equation}
which we express as a partial fraction decomposition 
\begin{equation}
    F_n(s) = (-i)^n\prod_{k=0}^n\frac{1}{s+i\Delta_k+\Theta_k} = \sum_{l=1}^r\sum_{p=1}^{m_l}\frac{C_{lp}}{(s+i\Delta_l+\Theta_l)^p}.
\end{equation}
The $(\Delta_l - i\Theta_l)$s are the $r$ distinct regularized cumulative detunings, each with multiplicity $m_l$. The cumulative regularization terms $\Theta_l$ are defined as $\sum_{i=0}^{l}\theta_i$, where the $\theta_i$ are the individual regularization terms attached to each detuning. Since all $\theta_i$ are strictly positive, one has $\Theta_l \neq \Theta_k$ for $k \neq l$, and thus all regularized cumulative detunings are non-degenerate, \textit{i.e.} $m_l = 1$. Therefore, 
\begin{equation}
    F_n(s) = \sum_{l=0}^n\frac{C_l}{(s+i\Delta_l+\Theta_l)}, \;\; C_l = \prod_{\substack{k=0 \\ k\neq l}}^n\frac{1}{\Delta_l-\Delta_k+i(\Theta_k-\Theta_l)},
\end{equation}
and consequently 
\begin{equation}
\label{Appendix Equation general expression of the time-dependent loop diagram}
   v_n(t) =
   (-1)^{N_L} \sum_{l=0}^n e^{-i\Delta_l t}e^{-\Theta_l t} \prod_{\substack{k=0 \\ k\neq l}}^n\frac{1}{\Delta_l-\Delta_k+i(\Theta_k-\Theta_l)}.
\end{equation}
The total time-dependent weight $W_n(t)$ of a given $n$th-order JLM transition operator is obtained by averaging the weights $w_{n}(t)$ over the possible perturbation time orderings and the $(n+1)$ choice of zeroth-order JLM operators, as illustrated next at first and second order.
\subsubsection*{First-order time weight computation}
\label{Appendix First-order time weight computation}
Consider the first-order contribution of a JLM transition operator $\hat \xi_i$ expansion due to the action of another JLM operator $\hat \xi_j$ acting on its left, thus leading to $\hat \xi_j \hat \xi_i$. The time-dependent part of the weight $w_1(t)$ of the corresponding first-order JLM diagram is given by Eq.~\eqref{Appendix Equation general expression of the time-dependent loop diagram}
\begin{equation}
    v_1(t) = \frac{-\left(e^{-i(\Delta_0-i\Theta_0)t}-e^{-i(\Delta_1-i\Theta_1)t} \right)}{\Delta_0-\Delta_1+i(\Theta_1-\Theta_0)}.
\end{equation}
Here $\Delta_0-i\Theta_0$ equals the regularized detuning $\delta_i-i\theta_i$ of the zeroth-order JLM transition operator, while
$\Delta_1-i\Theta_1=\delta_i+\delta_j-i(\theta_i+\theta_j)$, with $\delta_j-i\theta_j$ the regularized detuning associated to the first-order JLM transition operator. Due to the perturbation-time ordering, the two-photon JLM transition $\hat\xi_j\hat\xi_i$ -- with $\hat\xi_i$ and $\hat\xi_j$ denoting zeroth-order, one-photon JLM operators -- acquires a total weight $W_1(t)$ given by the average of the two possible $w_1(t)$ contributions (depending on which operator is perturbed, either $\hat \xi_i$ or $\hat \xi_j$). Hence, for a first-order JLM diagram producing the transition $\hat\xi_j\hat\xi_i$, the total time-dependent part $V_1(t)$ of $W_1(t)$ reads
\begin{equation}
\label{Appendix Equation total time-dependent weight for n=1}
    V_1(t) =\frac{1}{2}\bigg( \left[\frac{e^{-i\delta_it}e^{-\theta_it}}{\delta_j-i\theta_j}- \frac{e^{-i\delta_jt}e^{-i\theta_j t}}{\delta_i-i\theta_i}  \right]
    +e^{-i(\delta_i+\delta_j)t}e^{-(\theta_i+\theta_j)t}\left[\frac{1}{\delta_i-i\theta_i}-\frac{1}{\delta_j-i\theta_j}\right]\bigg),
\end{equation}
where the extra factor $-1$ arises from the permutation $\hat\xi_i\leftrightarrow\hat\xi_j$, which reverses operator order and flips the sign of that contribution because of the commutator with the interaction Hamiltonian. The weight of $\hat\xi_i\hat\xi_j$, when nonzero, is obtained by cyclic permutation of $\hat\xi_j\hat\xi_i$ and equals $-W_1(t)$. For a discrete spectrum in the dispersive regime one has $\delta_i\neq 0$ and $\delta_j\neq 0$, so one may safely take the limits $\theta_i\rightarrow 0$ and $\theta_j \rightarrow 0$ in Eq.~\eqref{Appendix Equation total time-dependent weight for n=1}, yielding
\begin{equation}
\label{Appendix Equation total time-dependent weight for n=1 and discrete case}
    V_1(t) = \frac{1}{2}\bigg( \left[\frac{e^{-i\delta_it}}{\delta_j}- \frac{e^{-i\delta_jt}}{\delta_i}  \right]
    +e^{-i(\delta_i+\delta_j)t}\left[\frac{1}{\delta_i}-\frac{1}{\delta_j}\right]\bigg).
\end{equation}
Over a time $T$ satisfying $T\gg 1/|\delta_i|$ and $T\gg 1/|\delta_j|$, $V_1(t)$ simplifies to
\begin{equation}
\label{Appendix Equation total time-dependent weight for n=1 after coarse-grained time}
    V_1(t) \approx \frac{1}{2}e^{-i(\delta_i+\delta_j)t}\left[\frac{1}{\delta_i}-\frac{1}{\delta_j}\right].
\end{equation}
Adding the coupling term 
\begin{equation}
    W_1(t) = \frac{\lambda^2}{2}e^{-i(\delta_i+\delta_j)t}\left[\frac{1}{\delta_i}-\frac{1}{\delta_j}\right].
\end{equation}
\subsubsection*{Second-order time weight computation}
\label{Appendix Second-order time weight computation}
At second order, a JLM transition operator $\hat\xi_k\hat\xi_j\hat\xi_i$, with $\hat\xi_s$ ($s=i,j,k$) denoting zeroth-order, one-photon transition operators, can be constructed by treating each of the three $\hat\xi_s$ in turn as the perturbed operator. Its canonical time weight -- where the time ordering is identical to the process ordering -- reads 
\begin{equation}
\label{Appendix Equation appendix n=2 time-weight without detuning degeneracies}
\begin{split}
    v_2^{\text{\tiny{canonical}}}(t) &= \frac{e^{-i(\Delta_0-i\Theta_0)t}}{\left(\Delta_0-\Delta_1+i(\Theta_1-\Theta_0)\right)\left(\Delta_0-\Delta_2+i(\Theta_2-\Theta_0)\right)} \\
    &+ \frac{e^{-(i\Delta_1-i\Theta_1) t}}{\left(\Delta_1-\Delta_0 +i(\Theta_0-\Theta_1)\right)\left(\Delta_1-\Delta_2+i(\Theta_2-\Theta_1)\right)} \\
    &+\frac{e^{-i(\Delta_2-i\Theta_2)t}}{\left(\Delta_2-\Delta_0+i(\Theta_0-\Theta_2)\right)\left(\Delta_2-\Delta_1+i(\Theta_1-\Theta_2)\right)},
\end{split}
\end{equation}
where $\Delta_l$ and $\Theta_l$ are the cumulative detunings and regularization terms at order $l$. For an ordering in which $\hat\xi_i$ is the zeroth-order operator, the canonical weight becomes
\begin{equation}
    v_2^{\text{\tiny{canonical}}}(t) = \frac{e^{-i(\delta_i-i\theta_i)t}}{\left(\delta_j-i\theta_j\right)(\delta_j+\delta_k-i(\theta_i+\theta_j))}
    - \frac{e^{-i\left(\delta_i+\delta_j -i(\theta_i+\theta_j)\right)t}}{\left(\delta_j-i\theta_j\right)\left(\delta_k-i\theta_k\right)}
    +\frac{e^{-i\left(\delta_i+\delta_j+\delta_k-i(\theta_i+\theta_j+\theta_k)\right)t}}{\left(\delta_j+\delta_k-i(\theta_j+\theta_k)\right)\left(\delta_k-i\theta_k\right)}.
\end{equation}
However, any of $\hat\xi_i,\hat\xi_j,\hat\xi_k$ can serve as the perturbed zeroth-order operator. For instance, if $\hat\xi_k$ is the perturbed operator (two adjacent transpositions relative to the canonical ordering), the weight reads
\begin{equation}
    v_2^{\text{\tiny{two adj. transp.}}}(t) = \frac{e^{-i(\delta_k-i\theta_k)t}}{\left(\delta_j-i\theta_j\right)(\delta_j+\delta_i-i(\theta_k+\theta_j))}
    - \frac{e^{-i\left(\delta_k+\delta_j -i(\theta_k+\theta_j)\right)t}}{\left(\delta_j-i\theta_j\right)\left(\delta_i-i\theta_i\right)}
    +\frac{e^{-i\left(\delta_i+\delta_j+\delta_k-i(\theta_i+\theta_j+\theta_k)\right)t}}{\left(\delta_j+\delta_i-i(\theta_j+\theta_i)\right)\left(\delta_i-i\theta_i\right)},
\end{equation}
which is equivalent to the permutation $i\leftrightarrow k$ of the canonical expression. If instead $\hat\xi_j$ is the zeroth-order perturbed operator, an extra minus sign appears because $\hat \xi_k$ acts on the left while $\hat \xi_i$ acts on the right with the commutator with the interaction Hamiltonian. Two additional terms should be added depending on whether $\hat \xi_k$ or $\hat \xi_i$ acts first on $\hat \xi_j$
\begin{equation}
    v_2^{\text{\tiny{one adj. transp.}}}(t) = \frac{-e^{-i\left(\delta_j -i\theta_j\right)t}+e^{-i\left(\delta_i+\delta_j -i(\theta_i-\theta_j)\right)t}+e^{-i\left(\delta_k+\delta_j-i(\theta_k+\theta_j)\right)t}-e^{-i\left(\delta_i+\delta_j+\delta_k -i(\theta_i+\theta_j+\theta_k)\right)t}}{\left(\delta_i-i\theta_i\right)\left(\delta_k-i\theta_k\right)}.
\end{equation}
The total second-order time weight is the average of these three contributions
\begin{equation}
\label{Appendix Equation total second order time weight}
    V_2(t) = \frac{1}{3}\left(v_2^{\text{\tiny{canonical}}}(t) +  v_2^{\text{\tiny{two adj. transp.}}}(t) + v_2^{\text{\tiny{one adj. transp.}}}(t)\right).
\end{equation}
Taking the regularization parameters to zero yields the following expression, which is valid in the absence of detuning degeneracies,
\begin{equation}
\label{Appendix Equation appendix n=2 total time-weight with no degeneracies}
\begin{split}
    V_2(t) =& \frac{1}{3}\bigg[\frac{e^{-i\delta_i t}}{\delta_j(\delta_j+\delta_k)}+\frac{e^{-i\delta_k t}}{\delta_j(\delta_i+\delta_j)}-\frac{e^{-i\delta_j t}}{\delta_i\delta_k} +\frac{e^{-i(\delta_i+\delta_j)t}}{\delta_k}\left(\frac{1}{\delta_i}-\frac{1}{\delta_j} \right) + \frac{e^{-i(\delta_k+\delta_j)t}}{\delta_i}\left(\frac{1}{\delta_k}-\frac{1}{\delta_j} \right) \\ 
    &+e^{-i(\delta_i+\delta_j+\delta_k)t}\left(\frac{1}{\delta_k(\delta_j+\delta_k)}+\frac{1}{\delta_i(\delta_i+\delta_j)} -\frac{1}{\delta_i\delta_k}\right)\bigg].
\end{split}
\end{equation}
The expression above assumes no detuning degeneracies at any stage (\textit{i.e}., conditions such as $\delta_i\neq -\delta_j$ and $\delta_j\neq -\delta_k$), since such degeneracies correspond to poles of $V_2(t)$ in Eq.~\eqref{Appendix Equation appendix n=2 total time-weight with no degeneracies}. This is discussed in the companion article~\cite{CompanionarticlePRA} where we argue that detuning degeneracies amount to renormalization of the free Hamiltonian which can be discarded. Eq.~\eqref{Appendix Equation total second order time weight} and Eq.~\eqref{Appendix Equation appendix n=2 total time-weight with no degeneracies} explicitly display the different time scales arising from one-, two-, and three-photon detunings along the JLM diagram pathways. If only the three-photon process is resonant, the expression simplifies for times long compared to all other detunings to
\begin{equation}
\label{Appendix Equation second-order time weight}
    V_2(t) = \frac{1}{3}e^{-i(\delta_i+\delta_j+\delta_k)t}\left(\frac{1}{\delta_k(\delta_k+\delta_j)}+\frac{1}{\delta_i(\delta_i+\delta_j)} -\frac{1}{\delta_i\delta_k}\right),
\end{equation}
valid over a time large compared with the inverse magnitudes of the other detunings. Thereupon adding the coupling term 
\begin{equation}
    W_2(t) = \frac{\lambda^3}{3}e^{-i(\delta_i+\delta_j+\delta_k)t}\left(\frac{1}{\delta_k(\delta_k+\delta_j)}+\frac{1}{\delta_i(\delta_i+\delta_j)} -\frac{1}{\delta_i\delta_k}\right).
\end{equation}
\subsection*{Three-photon resonance JLM diagrams, order by order}
In this section, we compute the correction to the Rabi interaction Hamiltonian $\hat H_{\text{int}}=\lambda \hat \sigma_x\otimes\left(\hat a_c+\hat a_c^\dagger\right)$ up to second-order, that is 
\begin{equation}
    \Delta \hat H_{\text{int}}^{(n\leq 2)} = \Delta \hat H_{\text{int}}^{(1)}(t)+\Delta \hat H_{\text{int}}^{(2)}(t).
\end{equation}
\subsubsection*{Zeroth order}
\begin{figure}
    \centering
    \includegraphics[width=0.7\linewidth]{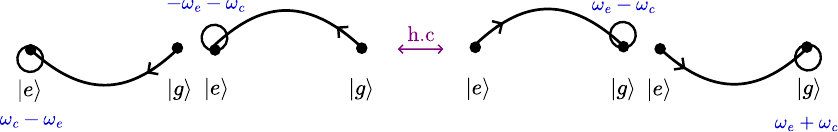}
    \caption{Zeroth-order diagram for the quantum Rabi interaction Hamiltonian. Detunings (in blue) are indicated for clarity, though this is not necessary in practice. }
    \label{Appendix figure zeroth-order JLM diagrams}
\end{figure}

At zeroth-order, the JLM transitions are the elements constitutive of $\hat H_{\text{int}}$, that is $\hat \sigma_+\otimes \hat a_c$, $\hat \sigma_- \otimes \hat a_c$ and their Hermitian conjugates; where we recall $\hat \sigma_+ = \ket{e}\bra{g}$ and $\hat \sigma_- = \ket{g}\bra{e}$. These operators have a graphical representation represented in Fig.~\ref{Appendix figure zeroth-order JLM diagrams}. In the Heisenberg picture, they have a time-dependent weight $W_0(t)$ equal to $e^{-i\delta t}$, where $\delta$ is the one-photon detuning corresponding to each transition. Over a time $T\gg 1/|\delta|$, these operators average out to zero and are thus discarded.
\subsubsection*{First order}
\label{Appendix First-order Rabi Hamiltonian}
\begin{figure}
    \centering
    \includegraphics[width=0.6\linewidth]{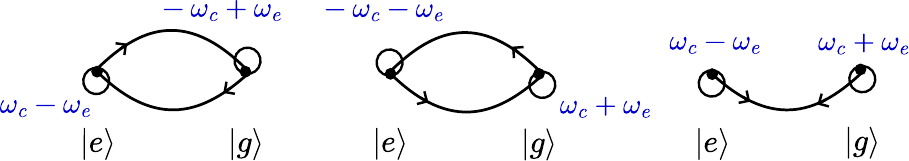}
    \caption{First-order diagrams for the quantum Rabi interaction Hamiltonian. Detunings (in blue) are indicated for clarity, though this is not necessary in practice.}
    \label{Appendix figure first-order JLM diagrams}
\end{figure}

At first order, the interaction Liouvillian $\mathcal L_{\text{int}}$ couples zeroth-order JLM transition operators together. We can identify two different topologies: (i) when the atomic system loops back to its initial state by emission and absorption events; (ii) when the atomic system travels back to its initial state by consecutive emission or absorption events, see Fig.~\ref{Appendix figure first-order JLM diagrams}. 
The weights $W_1(t)$ of each first-order JLM transition operator in $\Delta \hat H_{\text{int}}^{(1)}(t)$ have a time dependence $V_1(t)$ which may with Eq.~\eqref{Appendix Equation total time-dependent weight for n=1 and discrete case}
\begin{equation}
    V_1(t) = \frac{1}{2}\bigg( \left[\frac{e^{-i\delta_it}}{\delta_j}- \frac{e^{-i\delta_jt}}{\delta_i}  \right]
    +e^{-i(\delta_i+\delta_j)t}\left[\frac{1}{\delta_i}-\frac{1}{\delta_j}\right]\bigg)
\end{equation}
where $\delta_i$ and $\delta_j$ are, respectively, the one-photon detunings of the zeroth-order JLM transition operator $\hat \xi_i$ being perturbed and of the zeroth-order JLM transition operator $\hat \xi_j$ acting on $\hat \xi_i$ through $\mathcal L_{\text{int}}$. For the Stark-shift diagram in Fig.~\ref{Appendix figure first-order JLM diagrams}, we have for instance $\delta_i = \omega_c-\omega_e$ and $\delta_j = -\omega_c+\omega_e$, which in turn gives 
\begin{equation}
    V_{1}^{\text{Stark}}(t) = \frac{1}{2}\left(\left[\frac{e^{-i(\omega_c-\omega_e)t}}{-\omega_c+\omega_e}-\frac{e^{-i(-\omega_c+\omega_e)t}}{\omega_c-\omega_e}\right]+\left[\frac{1}{\omega_c-\omega_e}-\frac{1}{-\omega_c+\omega_e} \right] \right),
\end{equation}
for the first-order operator $\ket{g}\bra{g}\otimes \hat a_c^\dagger \hat a_c$. Discarding the fast-oscillating terms $e^{\pm i(\omega_c-\omega_e)t}$ over a time $T\gg 1/|\omega_c-\omega_e|$
\begin{equation}
    V_{1}^{\text{Stark}} = \frac{1}{\omega_c-\omega_e}.
\end{equation}
In a similar fashion, we have for $\ket{g}\bra{g}\otimes \hat a_c\hat a_c^\dagger$ and $\ket{g}\bra{g}\otimes \hat a_c \hat a_c$, respectively, $V_{1}^{\text{BS}} = -1/(\omega_c+\omega_e)$ and $V_{1}^{\text{off}}=0$ upon additionally discarding the $e^{-i2\omega_ct}$ fast-oscillating terms. $V_{1}^{\text{Stark}}$ and $V_{1}^{\text{BS}}$ correspond to the weight of the first-order JLM transition operator $\ket{g}\bra{g}\otimes \hat a_c^\dagger \hat a_c$ and $\ket{g}\bra{g}\otimes \hat a_c\hat a_c^\dagger$, respectively, where the perturbed zeroth-order operator is $\ket{e}\bra{g}\otimes \hat a_c$ and $\ket{e}\bra{g}\otimes \hat a_c^\dagger$, respectively. This amounts to reading the JLM diagrams in Fig.~\ref{Appendix figure first-order JLM diagrams} starting from the state $\ket{g}$. Starting from state $\ket{e}$ instead yields the first-order JLM transition operators $\ket{e}\bra{e}\otimes \hat a_c\hat a_c^\dagger$ and $\ket{e}\bra{e}\otimes \hat a_c^\dagger \hat a_c$ which have time dependent weights $-V_{1}^{\text{Stark}}$ and $-V_{1}^{\text{BS}}$, respectively. The first-order correction to the interaction Hamiltonian thus reads 
\begin{equation}
    \Delta \hat H_{\text{int}}^{(1)} = \lambda^2\left(V_{1}^{\text{Stark}}\left[\ket{g}\bra{g}\otimes \hat a_c^\dagger \hat a_c-\ket{e}\bra{e}\otimes \hat a_c\hat a_c^\dagger \right] +V_{1}^{\text{BS}}\left[\ket{g}\bra{g}\otimes \hat a_c \hat a_c^\dagger-\ket{e}\bra{e}\otimes \hat a_c^\dagger\hat a_c \right] \right).
\end{equation}
Using $\ket{g}\bra{g}+\ket{e}\bra{e} = 1$, $[\hat a_c,\hat a_c^\dagger] = 1$ and $\ket{g}\bra{g} = \left(1-\hat \sigma_z \right)/2$, $\ket{e}\bra{e}=\left(1+\hat \sigma_z \right)/2$ and discarding the constant energy terms
\begin{equation}
    \Delta \hat H^{(1)}_{\text{int}} = \left(\frac{\lambda^2}{\omega_e-\omega_c}+\frac{\lambda^2}{\omega_e+\omega_c}\right)\left(\hat \sigma_z \hat n_c+\frac{\hat \sigma_z}{2} \right),
\end{equation}
where we defined $\hat n_c = \hat a_c^\dagger \hat a_c$. 
\subsubsection*{Second order}
\begin{figure}
    \centering
    \includegraphics[width=0.5\linewidth]{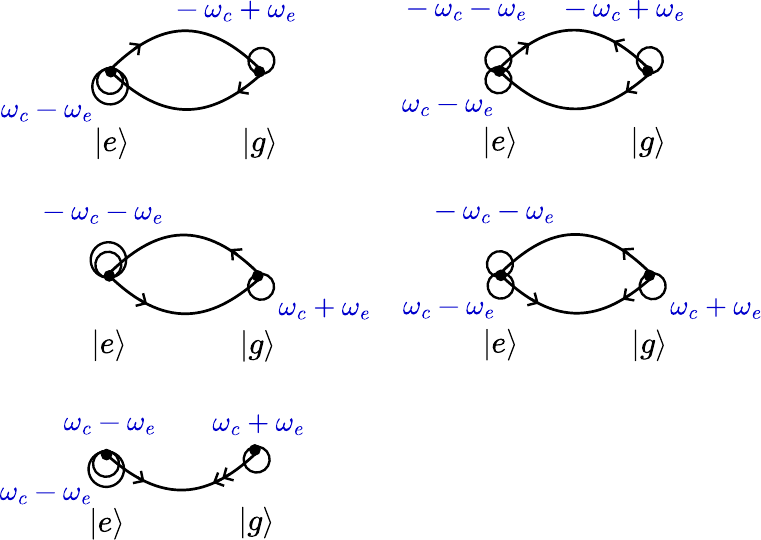}
    \caption{Second-order diagram for the quantum Rabi interaction Hamiltonian. Two free-evolution loops on the same atomic state simply corresponds to having a given one-photon detuning twice. Detunings (in blue) are indicated for clarity, though this is not necessary in practice.}
    \label{Appendix figure second-order JLM diagrams}
\end{figure}

At second order of the perturbation, five additional JLM diagrams may be constructed from the first-order JLM diagrams~\ref{Appendix figure first-order JLM diagrams}, and which we sketch in Fig.~\ref{Appendix figure second-order JLM diagrams}. 
Out of the five second-order JLM diagrams, the only one that amounts to a resonant transition is the one corresponding to three consecutive photon absorption events, that is $\ket{e}\bra{g}\otimes (\hat a_c)^3$. Indeed, the detunings of its constitutive one-photon JLM transitions are $\delta_i = \omega_c-\omega_e$, $\delta_j = \omega_c+\omega_e$ and $\delta_k=\omega_c-\omega_e$, which together lead to a total cumulative detunings $3\omega_c-\omega_e \approx 0$ since $\omega_e\approx 3\omega_c$ here. In contrast, the other diagrams will evolve at one-photon detunings and therefore average out over long times. The weight in the second-order correction $\Delta \hat H_{\text{int}}^{(2)}(t)$ to the Hamiltonian may hence be computed using Eq.~\eqref{Appendix Equation second-order time weight} as 
\begin{equation}
    W_2(t) = -\frac{9\lambda^3}{4\omega_e^2}e^{-i(3\omega_c-\omega_e)t},
\end{equation}
that is with $\omega_e \approx 3\omega_c$
\begin{equation}
    \Delta \hat H_{\text{int}}^{(2)} = -\frac{9\lambda^3}{4\omega_e^2}\hat \sigma_+ \otimes (\hat a_c)^3 +\text{h.c}.
\end{equation}
\subsection*{Intrinsic Rabi frequency and Rabi oscillations in operator space : detailed derivations}
We show how the polaritonic joint population term $\hat \sigma_z\hat n_c$ appears in both the resonant and dispersive regimes. In the resonant regime, it contributes as a renormalization of the coupling term $\lambda$, while in the dispersive regime, it manifests as a renormalization of the detuning. This provides a unified picture of the manifestation of light-matter hybridization across these two regimes. 
\subsubsection{Intrinsic Rabi frequency}
Consider the Jaynes-Cummings Hamiltonian of the main text $\hat H_{\text{JC}} = \frac{\omega_e}{2}\hat \sigma_z + \omega_c \hat n_c+ \lambda\left(\hat \sigma_+ \hat a_c+\hat \sigma_- \hat a_c^\dagger\right)$. Under the adjoint Liouvillian $\mathcal L_{\text{JC}}(*) = -i[*,\hat H_{\text{JC}}]$, the time evolution of the four-dimensional vector $X' = (\hat n_c, \hat \sigma_z,\hat \sigma_+\hat a_c, \hat \sigma_-\hat a_c^\dagger)^T$ may be expressed as $\dot{X}' = -iM'X'$ where 
\begin{equation}
    M' = \begin{bmatrix}0 & 0 & 2\lambda & -2\lambda \\
    0 & 0 & -\lambda &\lambda \\
    \lambda/2 & 0 &\Delta & 0 \\
    -\lambda/2 & 0 & 0 &-\Delta \end{bmatrix},
\end{equation}
where $\Delta = \omega_c-\omega_e$ is the one-photon detuning associated to the light-matter transition $\hat \sigma_+\hat a_c$. We derived the time-dynamics equation of $X'$ by using the commutation relation $[\hat a_c,\hat a_c^\dagger] = \hat{\mathbb{1}}_c$ and by discarding irrelevant identity operators. The eigenvalues of $M$ can be computed as $\text{Sp}(M)=\{0,0,\pm \Omega'\}$ where $\Omega' = \sqrt{\Delta^2+2\lambda^2}$, which does not match the photon-number-dependent Rabi frequency in the state-centric framework upon restricting the Jaynes-Cummings Hamiltonian to the $\{\ket{e,n},\ket{g,n+1}\}_{n\geq 0}$ manifold. \\
If instead we consider the four-dimensional vector $X=(\hat n_c, \hat \pi,\hat \sigma_+ \hat a_c, \hat \sigma_-\hat a_c^\dagger)^T$, with $\hat \sigma_z \rightarrow \hat \pi = \hat \sigma_z/2 + \hat \sigma_z \hat n_c$, where $\hat \sigma_z\hat n_c$ is the polaritonic joint population operator, the time-evolution reads $\dot{X} = -iMX$ with
\begin{equation}
    M = \begin{bmatrix}
    0 & 0 & -\lambda & \lambda \\ 
    0 & 0 & 2\lambda & -2\lambda \\ 
    0 & \lambda & \Delta & 0 \\ 
    0 & -\lambda & 0 & -\Delta
    \end{bmatrix},
\end{equation}
up to the first occurrence of a polaritonic joint population operator; high-order contributions are neglected \textit{i.e.}, we retain only the first additional operators generated by the action of the interaction Liouvillian on $\mathcal S' = \{\hat n_c, \hat \sigma_z, \hat \sigma_+ \hat a_c, \hat \sigma_- \hat a_c^\dagger\}$. The resulting eigenvalues are $\text{Sp}(M) = \{0,0,0, \pm \Omega\}$ with $\Omega = \sqrt{\Delta^2+4\lambda^2}$. 
\subsubsection*{Rabi oscillations in operator space}
Let us define $P_\Omega = M^2/\Omega^2$ the projector onto the eigenspaces with eigenvalues $\pm \Omega$, which removes static components and keeps only components at frequency $\Omega$. $P_\Omega$ can be calculated as 
\begin{equation}
    P_\Omega = \frac{1}{\Delta^2+4\lambda^2}\begin{bmatrix}
    0 & -2\lambda^2 & -\Delta\lambda & -\Delta\lambda \\
     0 & 4\lambda^2 & 2\Delta\lambda & 2\Delta\lambda \\
     0 & \Delta\lambda & \Delta^2+2\lambda^2 & -2\lambda^2 \\
     0 & \Delta\lambda & -2\lambda^2 & \Delta^2+2\lambda^2
    \end{bmatrix},
\end{equation}
In the resonant or near-resonant regime, $\lambda \gg |\Delta|$ and $P_\Omega$ simplifies to 
\begin{equation}
    P_\Omega \approx \begin{bmatrix} 
    0 & -1/2 & 0 & 0 \\ 
    0 & 1 & 0 & 0 \\ 
    0 & 0 & 1/2 & -1/2 \\ 
    0 & 0 & -1/2 & 1/2
    \end{bmatrix}.
\end{equation}
The total population sector, accounting for both the disjoint $\hat n_c$ and joint population polaritonic term $\hat \pi$, are decoupled from the transition sector $\{\hat \sigma_+\hat a_c, \hat \sigma_-\hat a_c^\dagger\}$. Furthermore, considering the symmetric and antisymmetric superpositions $\hat T_s = \hat \sigma_+\hat a_c+\hat \sigma_-\hat a_c^\dagger$ and $\hat T_a =\hat \sigma_+\hat a_c-\hat \sigma_-\hat a_c^\dagger$, one observe that $\hat T_s$ is projected out while $\hat T_a$ survives and oscillates. In analogy with the Bloch vector, this may be understood as follows. At resonance, the symmetric transition operator becomes a constant of motion and defines a rotation axis of the dynamics. The oscillatory subspace is spanned by the antisymmetric transition operator and the joint population operator whereas the photon-number operator is not an independent degree of freedom but is slaved to the joint population operator. \\
Equivalently, in the dispersive regime $|\Delta| \gg \lambda$, the projector $P_\Omega$ approximates to 
\begin{equation}
     P_\Omega \approx \begin{bmatrix} 
   0 & 0 & 0 & 0 \\
   0 & 0 & 0 & 0 \\
   0 & 0 & 1 & 0 \\
   0 & 0 & 0 & 1 
    \end{bmatrix},
\end{equation}
effectively decoupling the population and transition sectors and the elements of the latter. In the dispersive regime, the rotation axis is defined along the population operator. As a result, the population sector becomes stationary, while the transition operators span the oscillatory subspace. In contrast to the resonant regime, no destructive interference occurs within the transition sector. \\
Hence, the resonant and dispersive regimes correspond to rotations about orthogonal axes in operator space. In the resonant case, the symmetric transition operator defines a rotation axis and is frozen, whereas in the dispersive regime the population operator freezes and defines a rotation axis as the transition sector contains the Rabi oscillations. The oscillatory subspace is therefore rotated from the population sector to the transition sector as the system crosses from resonance to the dispersive regime.
\bibliography{ref_JLM_dia}